# Large-scale integration of near-indistinguishable artificial atoms in hybrid photonic circuits

Noel H. Wan[1,*,a], Tsung-Ju Lu[1,*,b], Kevin C. Chen[1], Michael P. Walsh[1], Matthew E. Trusheim[1], Lorenzo De Santis[1], Eric A. Bersin[1], Isaac B. Harris[1], Sara L. Mouradian[1,2], Ian R. Christen[1], Edward S. Bielejec[3], and Dirk Englund[1,c]

[1] Research Laboratory of Electronics, MIT, Cambridge, MA 02139, USA
[2] Present address: University of California Berkeley, California 94720, USA
[3] Sandia National Laboratories, Albuquerque, New Mexico 87185, USA
* These authors contributed equally to this work

**Abstract**

A central challenge in developing quantum computers and long-range quantum networks lies in the distribution of entanglement across many individually controllable qubits[1]. Colour centres in diamond have emerged as leading solid-state 'artificial atom' qubits[2,3], enabling on-demand remote entanglement[4], coherent control of over 10 ancillae qubits with minute-long coherence times[5], and memory-enhanced quantum communication[6]. A critical next step is to integrate large numbers of artificial atoms with photonic architectures to enable large-scale quantum information processing systems. To date, these efforts have been stymied by qubit inhomogeneities, low device yield, and complex device requirements. Here, we introduce a process for the high-yield heterogeneous integration of 'quantum micro-chiplets' (QMCs) – diamond waveguide arrays containing highly coherent colour centres – with an aluminium nitride (AlN) photonic integrated circuit (PIC). Our process enables the development of a 72-channel defect-free array of germanium-vacancy (GeV) and silicon-vacancy (SiV) colour centres in a PIC. Photoluminescence spectroscopy reveals long-term stable and narrow average optical linewidths of 54 MHz (146 MHz) for GeV (SiV) emitters, close to the lifetime-limited linewidth of 32 MHz (93 MHz). Additionally, inhomogeneities in the individual qubits can be compensated *in situ* with integrated tuning of the optical frequencies over 100 GHz. The ability to assemble large numbers of nearly indistinguishable artificial atoms into phase-stable PICs provides an architecture toward multiplexed quantum repeaters[7,8] and general-purpose quantum computers[9–11].

**Main text**

Artificial atom qubits in diamond combine minute-scale quantum memory times[5] with efficient spin-photon interfaces[2], making them attractive for processing and distributing quantum information[1,3]. However, the low device yield of functional qubit systems presents a critical barrier to large-scale quantum information processing (QIP). Furthermore, although individual diamond cavity systems coupled to artificial atoms can now achieve excellent performance, the lack of active chip-integrated photonic components and wafer-scale single crystal diamond currently prohibit scaling to large-scale QIP applications[8–11]. A promising method to alleviate these constraints is heterogeneous integration (HI), which is increasingly used in advanced microelectronics to assemble separately fabricated sub-components into a single, multifunctional chip. HI approaches have also recently been used to integrate PICs with quantum devices, including quantum dot single-photon sources[12,13], superconducting nanowire single-photon detectors[14], and nitrogen-vacancy (NV) centre diamond waveguides[15]. However, these demonstrations assembled components one-by-one, which presents a formidable scaling challenge. The diamond 'quantum micro-chiplet (QMC)' introduced here significantly improves HI assembly yield and accuracy to enable a 72-channel defect-free waveguide-coupled artificial atoms-photonics microchip. The PIC features diamond emitters with high coupling efficiencies, optical coherences near the lifetime limit, and integrated control to compensate for spectral inhomogeneities on chip.

Correspondence: [a]noelwan@mit.edu, [b]tsungjul@mit.edu, [c]englund@mit.edu

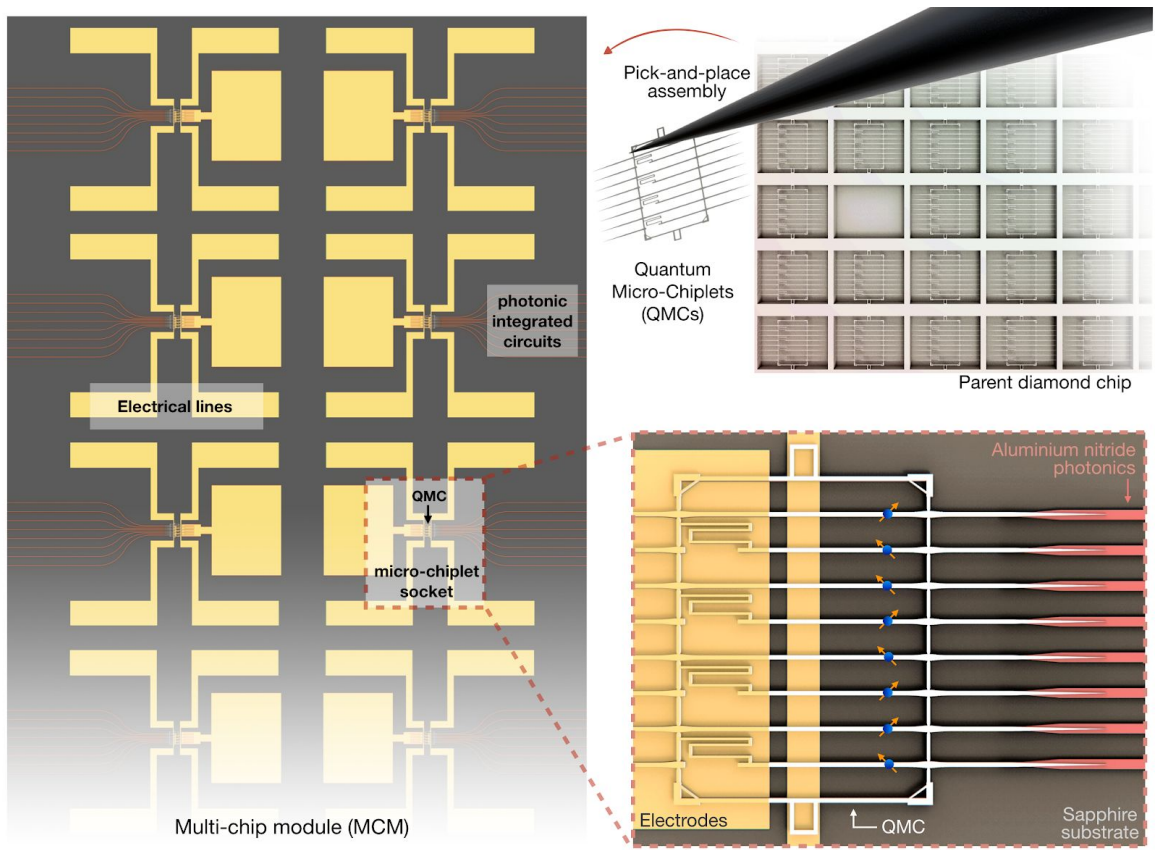

**Figure 1| Scalable integration of artificial atoms with photonics.** The separate fabrication of sub-components before their final assembly maximises the yield, size, and performance of the hybrid emitter-photonics chip. A pick-and-place method transfers pre-screened QMCs from their parent diamond chip into a socket containing efficient photonic interfaces, as well as electrical wires for controlling colour centres.

Figure 1 illustrates the QMC-PIC assembly process. The multi-chip module contains waveguides in a single-crystal AlN layer that has low autofluorescence and loss in the visible spectrum[16,17]. Additionally, it contains electrical wires for controlling colour centres. As detailed below, we pre-screened and subsequently transferred suitable QMCs into the micro-chiplet sockets using a pick-and-place process.

First, we demonstrate the fabrication and high-yield coupling of emitters into a diamond QMC. We chose the negatively charged GeV and SiV centres because of their stable optical and spin properties[18–23] in nanophotonic devices. The process begins with focused ion beam (FIB) implantation of $Ge^+$ and $Si^+$ into a 1 μm pitch square array in a single-crystal diamond substrate, followed by high temperature annealing (see Supplementary Information)[24]. This process generates spots of tightly localised GeV centres (depth of ~74 nm, vertical straggle ~12 nm, lateral FWHM distribution ~40 nm) and SiV centres (~113 nm, ~19 nm, ~50 nm), which we then registered relative to pre-fabricated alignment markers by photoluminescence (PL) microscopy. We fabricated the QMCs over the emitter arrays using a combination of electron-beam lithography (EBL) and quasi-isotropic etching[25,26]. Fig. 2a shows scanning electron micrographs (SEMs) of various suspended chiplets containing 8- or 16-channel waveguide arrays connected by diamond 'trusses', as seen in the close-up SEMs in Fig. 2b,c and Fig. 2g, respectively. Structurally, much larger arrays are fabricable and integrable: we successfully transferred QMCs with as many as 64 waveguide components (see Supplementary Information). Despite a misalignment between the FIB mask and the QMC patterns, the PL scans showed that 40% of 8-channel QMCs are "defect-free" (i.e. they have more than one stable colour centre per waveguide) as shown in Fig. 2e (see Supplementary Information). The defect-free yield of the 16-channel QMCs was lower as these are more susceptible to misalignment, so we did not use them in this
2

study. With improvements in FIB alignment and lithography, much higher yield should be possible in future work.

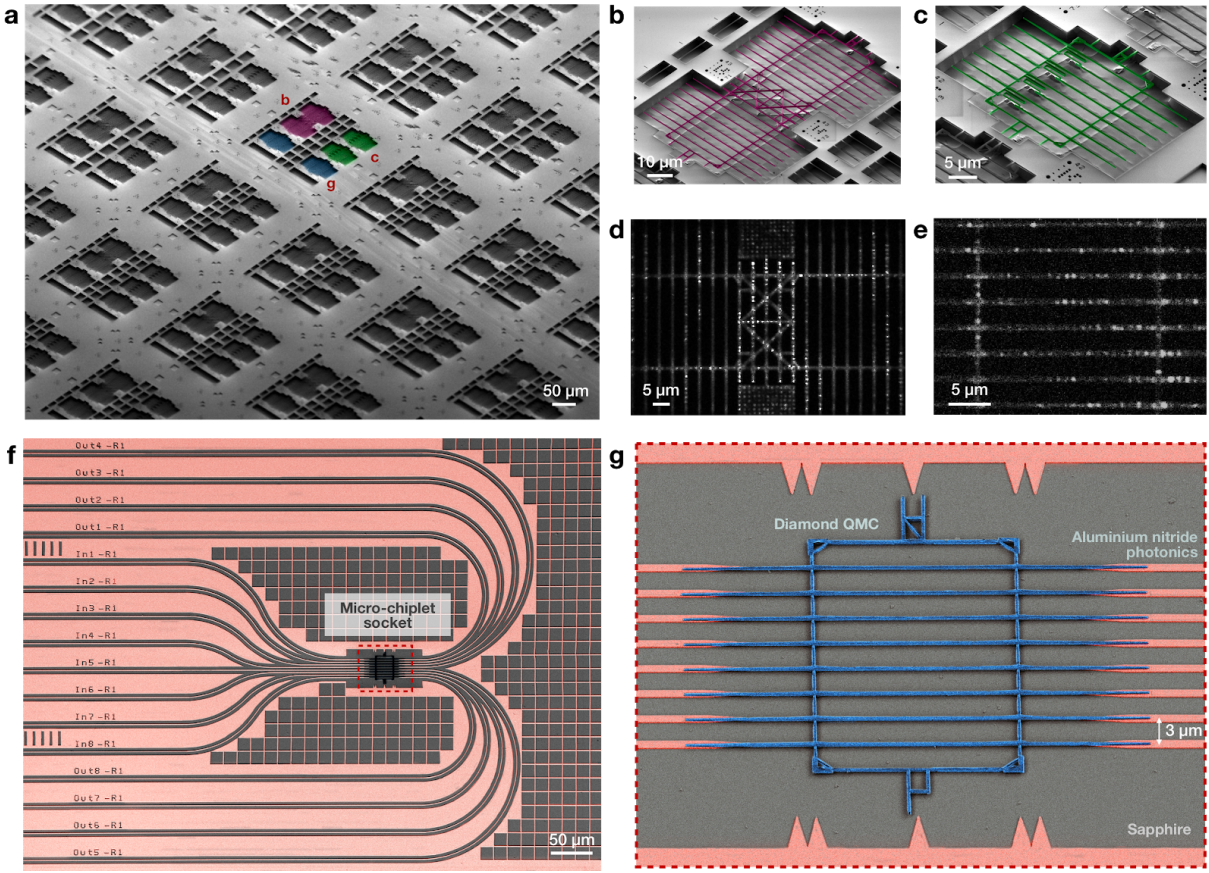

**Figure 2 | Fabrication and integration of QMC with integrated photonics. a,** SEM overview of the parent diamond chip containing over 500 micro-chiplets for heterogeneous integration. **b,** A 16-channel QMC. **c,** An 8-channel QMC with varying mechanical beam rigidity. **d,** PL map of GeV centres (bright spots) in a 16-channel QMC. **e**, PL map of SiV centres (bright spots) in a defect-free 8-channel QMC. **f,** An AlN-on-sapphire integrated photonics module that interfaces with the diamond QMC placed in the chiplet socket. **g**, Close-up SEM of the diamond QMC and AlN photonic interfaces.

Figure 2f shows one of ten micro-chiplet sockets connecting 8 input and 8 output waveguide arrays to an 8-channel QMC. We fabricated this PIC on a wafer of single-crystal AlN on a sapphire substrate using EBL and chlorine reactive ion etching[16] (see Supplementary Information). Using piezo-controlled micro-manipulators, we were able to transfer QMCs into the micro-chiplet sockets with a success rate of 90%. The diamond waveguides (width 340 nm, height 200 nm) transfer light into the AlN waveguides (width 800 nm, height 200 nm) through inverse tapered sections with simulated efficiency of 97% (98%) at 602 nm (737 nm) wavelength. The SEM of an assembled device in Fig. 2g shows a transverse placement error of (38 ± 16) nm. For such typical errors, simulations indicate a drop in coupling efficiency by 10% or 0.46 dB (see Supplementary Information). We find that the transfer of the QMCs is substantially easier than for individual waveguides due to its rigidity and many alignment features.

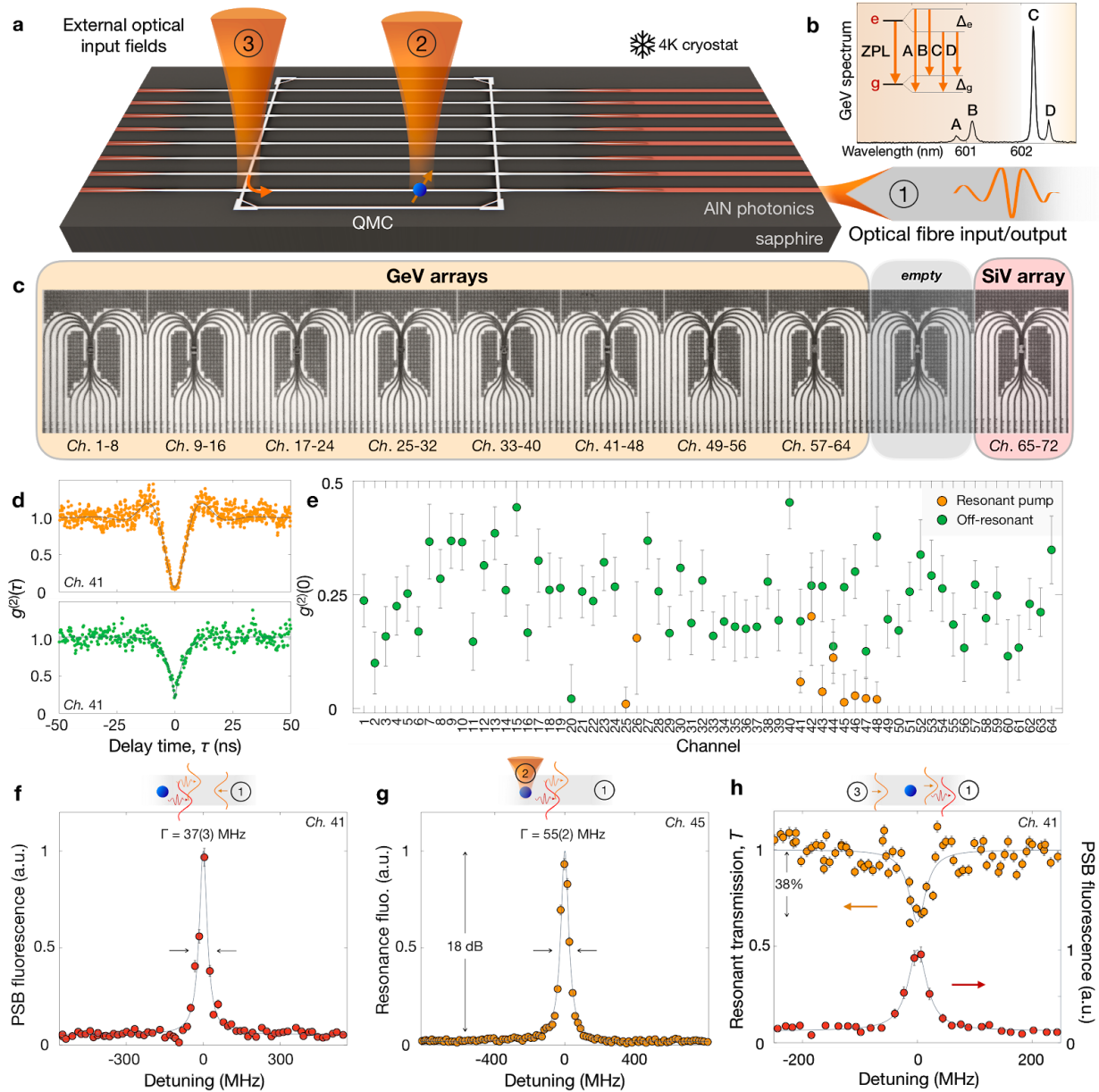

**Figure 3 | Integrated quantum photonics with colour centres. a,** Experimental setup in a 4K cryostat showing the input and output optical interfaces ①, ②, and ③. **b,** Energy level and spectrum of a GeV centre. Resonant excitation probed transition C, which is the brightest and narrowest line. **c,** Optical image of ten QMC-populated micro-chiplet sockets containing GeV or SiV centres. The 'empty' module indicates a failed QMC placement. **d,** Auto-correlation measurements of a single GeV in Channel 41 under off-resonant 2 mW, 532 nm excitation (bottom) and under resonant 10 nW, 602 nm excitation (top) **e**, Observation of waveguide-coupled single photons from every GeV channel in the PIC. **f,** Photoluminescence excitation (PLE) spectrum (FWHM linewidth $\Gamma = 37(3)$ MHz) of a single GeV in Channel 41 with all-fibre excitation and detection routed on-chip via ①. **g,** Excitation via ② and fluorescence detection via ①. This geometry allows GeV resonance fluorescence detection at least 18 dB above background, without filtering (spectral, temporal, polarisation). **h,** In transmission, a single GeV center causes coherent extinction of $\Delta T/T = 38(9)\%$ (orange curve, $\Gamma = 35(15)$ MHz); the PLE spectrum is shown in the red curve ($\Gamma = 40(5)$ MHz).

We performed experiments in a closed-cycle cryostat with a base temperature below 4K, as illustrated in Fig. 3a. The optical fibre labeled ① couples pump light (fluorescence) to (from) the QMC via the AlN waveguides. A microscope objective also provides optical access to the QMC, e.g., to a colour centre (optical interface ②) or a scattering site (③). Fig. 3b shows the energy level and emission spectrum of a single GeV when pumped through ② and collected through ①. Off-resonant excitation using 532 nm light with off-chip pump filtering in this configuration enables the rapid identification of single emitters (indicated by a photon autocorrelation function $g^{(2)}(0) < 0.5$). The bottom panel of Fig. 3d shows a typical photon antibunching ($g^{(2)}(0) = 0.19(7)$) from a single GeV centre (Channel 41) pumped near saturation, without background or detector jitter correction. Under the resonant excitation (10 nW, 602 nm) of transition C (see Fig. 3b) of the zero-phonon line (ZPL), the photon purity improves to $g^{(2)}(0) = 0.06(2)$ (top panel of Fig. 3d). By repeated measurements (pump through ② and collection through ①), we identified single GeV emitters in all integrated QMC waveguides as summarised by their photon statistics in Fig. 3e.

Next, we investigated the optical coherence of the GeV using all-fibre spectroscopy. Fig. 3f shows the photoluminescence excitation (PLE) spectrum of the Channel 41 GeV as we scanned a resonant laser across its ZPL (transition C) with both excitation and detection through the fibre interface ①. The measured linewidth of $\Gamma = \Gamma_0 + 2\Gamma_d = 37$ MHz (3 MHz fit uncertainty) is near the lifetime limit $\Gamma_0 = 1/2\pi\tau = 24(2)$ MHz, obtained from the excited state lifetime $\tau$ (see Supplementary Information).

The PIC geometry also enables the direct detection of ZPL resonance fluorescence without any spectral, temporal, or polarisation filtering, even under resonant excitation. Fig. 3g plots the resonance fluorescence obtained for top excitation (②) and waveguide collection without filtering in the detection via ①. By polarizing the pump $E$-field along the waveguide axis to minimize excitation of the TE waveguide mode, this cross-excitation/detection configuration achieves a ZPL intensity 18 dB above background, comparable to free-space diamond entanglement experiments using cross-polarisation and time-gated detection[27].

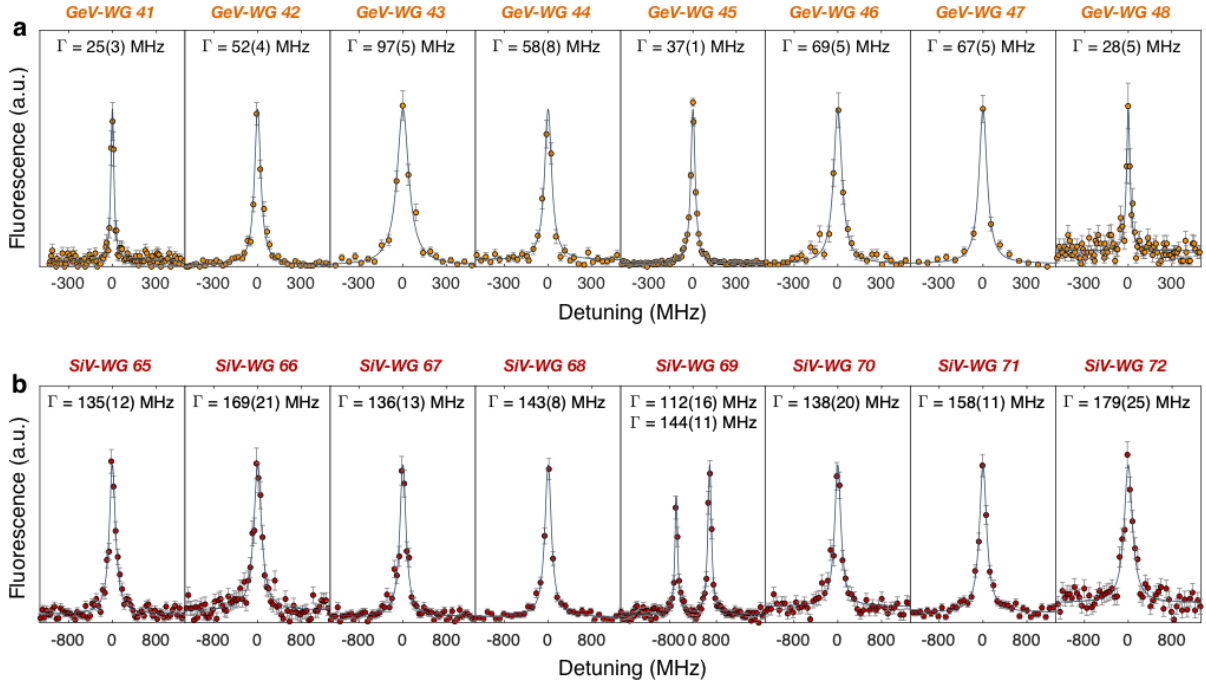

**Figure 4 | Defect-free arrays of optically coherent waveguide-coupled emitters. a,** PLE spectra of GeV centres in each waveguide of a characteristic 8-channel GeV QMC, with a mean (± standard deviation) linewidth of $\Gamma = 54$ ($\pm$ 24) MHz. **b,** PLE spectra of SiVs in an 8-channel SiV QMC, with a mean linewidth $\Gamma = 146$ ($\pm$ 20) MHz.

According to finite-difference time-domain simulations of our system, an ideal emitter in the optimal configuration has a spontaneous emission coupling efficiency of β = 0.8 into the diamond waveguide. Experimentally, we measured this efficiency by measuring the transmission of a laser field through a single GeV centre (Fig. 3h). By injecting a laser field through ③ and monitoring the transmission $T$ via ①, we observed an extinction of $1 − T = 0.38(9)$ when on resonance with the GeV centre. This extinction places a lower bound of the emitter-waveguide cooperativity at C = 0.27(10) and β = 0.21(6). By accounting for residual line broadening and for the ZPL emission fraction (~0.6), the dipole-waveguide coupling efficiency is at least 0.55(18); see Supplementary Information for other factors that reduce β.

The excellent coherence of the GeV centre in Channel 41 is not unique. Fig. 4a reports the linewidths of every channel in a characteristic 8-channel GeV diamond chiplet, all measured through the on-chip routing of fluorescence into an optical fibre. We find a mean ± standard deviation normalised linewidth of $\Gamma/\Gamma_0 = 1.7 ± 0.7$, with GeV channels 41, 45, and 48 exhibiting lifetime-limited values of 1.0(2), 0.9(1), and 1.0(2), respectively. From these measurements, we also obtained the inhomogeneous ZPL C transition frequency distribution of 85 GHz. In waveguides channels 65-72 (see Fig. 3c), we investigated an 8-channel QMC of SiV centres, whose linewidths are summarised in Fig. 4b. The SiV centres are also within a factor of $\Gamma/\Gamma_0 = 1.6 ± 0.2$ from SiV centres in bulk diamond[28], with an inhomogeneous distribution of 30 GHz. In all these measurements, we averaged each PLE spectrum over ~5 minutes (5000 experiments), demonstrating the long-term stability of the optical coherences in the heterogeneously integrated nanophotonic devices.

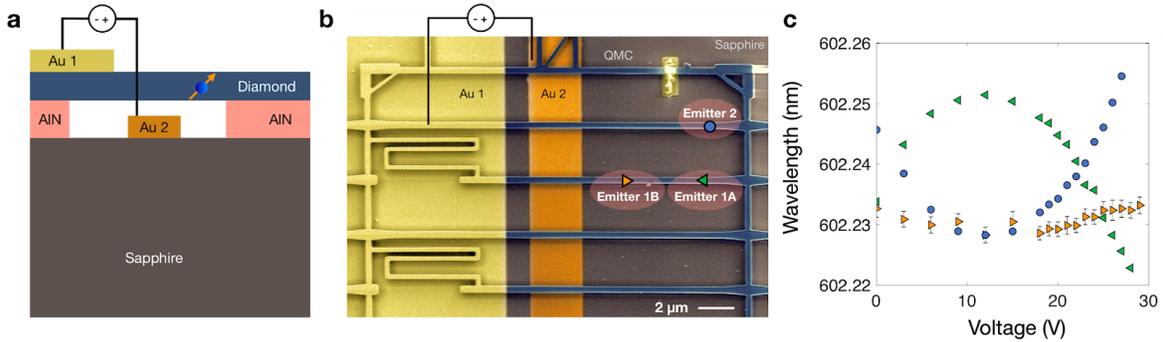

**Figure 5 | Controlling the optical transitions of colour centres on a PIC. a,** We applied a DC bias between the metal layer Au 1 on diamond and metal Au 2 on the substrate to electrostatically actuate the QMC. **b,** SEM of the gold electrodes and an integrated QMC with varying mechanical rigidity. In this experiment, we investigated the optical response of emitters 1A, 1B, and 2 to strain. **c,** Intra-waveguide Emitter 1A and Emitter 1B overlaps spectrally at 24.5 V. Inter-waveguide overlapping between Emitter 2 and Emitter 1A (1B) occurs at 2 V (12 V). Error bars for Emitter 1A and Emitter 2 are smaller than the data points.

Finally, we take advantage of the electrical layers in our chip to tune the optical transitions of diamond colour centres. As recently reported, the deformation of a diamond waveguide modifies the orbital structure of an embedded colour centre and its optical transition[29], making it possible to strain-tune separate emitters to the same frequency[30,31]. Here, we use a QMC that consists of waveguides with different lengths and therefore different strain responses. As shown in Fig. 5a, we fabricated a capacitive actuator consisting of one gold electrode (Au 1) on the diamond top side, separated transversely by 1.5 μm from a gold ground plane (Au 2) on the sapphire region of the AlN chip. Fig. 5b shows the fabricated chip. By applying DC voltages, as shown in Fig. 5c, the optical transition C of Emitter 1A (green) tunes to that of Emitter 1B (orange) near 24.5 V. At 2 V and 12 V, Emitter 2 (blue) in another waveguide channel also overlaps spectrally with Emitters 1A and 1B, respectively, both initially separated from Emitter 2 by ~10 GHz. At higher voltages, we observed tuning ranges up to 100 GHz, larger than the inhomogeneous centre frequency distribution of 88 GHz, and only limited by stiction between the QMC and the substrate (see Supplementary Information).

In conclusion, we have demonstrated an approach for high-yield integration of large numbers of waveguide-coupled, optically coherent diamond colour centres into photonic integrated circuits. Our approach for QMC integration into hybrid PICs would also enable large-scale assembly of other emitter types including NV centres[15], other diamond Group-IV quantum memories[32], quantum dots[33], and rare-earth ion dopants[34,35]. The same nanofabrication process[25,26] can produce diamond photonic crystal nanocavities with quality factors over $10^4$, which would enable atom-photon cooperativities in excess of 100. The inclusion of CMOS electronics in our modular architecture would also provide a path towards large-scale chip-integrated quantum control[36,37]. These advances, taken together with the deterministic assembly of large-scale artificial atoms with PICs with phase-stable interferometers[16,38] and modulators[39,40,41] in AlN or lithium niobate integrated photonics[42,43], set the stage for high-rate photon-mediated entanglement generation that underpins applications from multiplexed quantum repeaters[44,45] to modular quantum computing[9,11,46].

**Acknowledgements**

The focused ion beam implantation work was performed at the Center for Integrated Nanotechnologies, an Office of Science User Facility operated for the U.S. Department of Energy (DOE) Office of Science. Sandia National Laboratories is a multimission laboratory managed and operated by National Technology & Engineering Solutions of Sandia, LLC, a wholly owned subsidiary of Honeywell International, Inc., for the U.S. DOE's National Nuclear Security Administration under contract DE-NA-0003525. The views expressed in the article do not necessarily represent the views of the U.S. DOE or the United States Government. This work made use of the Shared Experimental Facilities supported in part by the MRSEC Program of the National Science Foundation under award number DMR - 1419807. We thank Daniel Perry for providing the focused ion beam implantation at Sandia National Laboratories, and Di Zhu and Cheng Peng for assistance with wire bonding. N.H.W. acknowledges support from the Army Research Laboratory Center for Distributed Quantum Information (CDQI). T.-J. L. acknowledges support from the Department of Defense National Defense Science and Engineering Graduate Fellowship as well as the Air Force Research Laboratory RITA program FA8750-16-2-0141. K.C.C. acknowledges funding support by the National Science Foundation (NSF) Graduate Research Fellowships Program and ARL CDQI. M.P.W. acknowledges support from the NSF Center for Integrated Materials. M.T. acknowledges support by an appointment to the Intelligence Community Postdoctoral Research Fellowship Program at the Massachusetts Institute of Technology, administered by the Oak Ridge Institute for Science and Education through an interagency agreement between the US Department of Energy (DOE) and the Office of the Director of National Intelligence. L.D.S acknowledges support from MIT Lincoln Laboratory. E.A.B. was supported by a NASA Space Technology Research Fellowship and the NSF Center for Ultracold Atoms. I.B.H is supported by the DOE 'Photonics at Thermodynamic Limits' Energy Frontier Research Center under grant DE-SC0019140**.** S.L.M was supported by the NSF EFRI ACQUIRE program.

**References**

1. Wehner, S., Elkouss, D. & Hanson, R. Quantum internet: A vision for the road ahead. *Science* **362**, (2018).

2. Awschalom, D. D., Hanson, R., Wrachtrup, J. & Zhou, B. B. Quantum technologies with optically interfaced solid-state spins. *Nature Photonics* **12**, 516–527 (2018).

3. Atatüre, M., Englund, D., Vamivakas, N., Lee, S.-Y. & Wrachtrup, J. Material platforms for spin-based

photonic quantum technologies. *Nature Reviews Materials* **3**, 38–51 (2018).

4. Humphreys, P. C. *et al.* Deterministic delivery of remote entanglement on a quantum network. *Nature* **558**, 268–273 (2018).

5. Bradley, C. E. *et al.* A Ten-Qubit Solid-State Spin Register with Quantum Memory up to One Minute. *Physical Review X* **9**, (2019).

6. Bhaskar, M. K. *et al.* Experimental demonstration of memory-enhanced quantum communication. *arXiv [quant-ph]* (2019).

7. Munro, W. J., Harrison, K. A., Stephens, A. M., Devitt, S. J. & Nemoto, K. From quantum multiplexing to high-performance quantum networking. *Nat. Photonics* **4**, 792 (2010).

8. Muralidharan, S. *et al.* Optimal architectures for long distance quantum communication. *Sci. Rep.* **6**, 20463 (2016).

9. Nemoto, K. *et al.* Photonic Architecture for Scalable Quantum Information Processing in Diamond. *Phys. Rev. X* **4**, 031022 (2014).

10. Monroe, C. *et al.* Large-scale modular quantum-computer architecture with atomic memory and photonic interconnects. *Phys. Rev. A* **89**, 022317 (2014).

11. Nickerson, N. H., Fitzsimons, J. F. & Benjamin, S. C. Freely Scalable Quantum Technologies Using Cells of 5-to-50 Qubits with Very Lossy and Noisy Photonic Links. *Phys. Rev. X* **4**, 041041 (2014).

12. Kim, J.-H. *et al.* Hybrid integration of solid-state quantum emitters on a silicon photonic chip. *Nano Lett.* **17**, 7394–7400 (2017).

13. Elshaari, A. W. *et al.* On-chip single photon filtering and multiplexing in hybrid quantum photonic circuits. *Nat. Commun.* **8**, 379 (2017).

14. Najafi, F. *et al.* On-chip detection of non-classical light by scalable integration of single-photon detectors. *Nat. Commun.* **6**, (2015).

15. Mouradian, S. L. *et al.* Scalable Integration of Long-Lived Quantum Memories into a Photonic Circuit. *Phys. Rev. X* **5**, 031009 (2015).

16. Lu, T.-J. *et al.* Aluminum nitride integrated photonics platform for the ultraviolet to visible spectrum. *Opt. Express* **26**, 11147–11160 (2018).

17. Liu, X. *et al.* Ultra-high-Q UV microring resonators based on a single-crystalline AlN platform. *Optica* **5**, 1279–1282 (2018).

18. Sukachev, D. D. *et al.* Silicon-Vacancy Spin Qubit in Diamond: A Quantum Memory Exceeding 10 ms with Single-Shot State Readout. *Phys. Rev. Lett.* **119**, 223602 (2017).

19. Becker, J. N. *et al.* All-Optical Control of the Silicon-Vacancy Spin in Diamond at Millikelvin Temperatures. *Phys. Rev. Lett.* **120**, 053603 (2018).

20. Bhaskar, M. K. *et al.* Quantum Nonlinear Optics with a Germanium-Vacancy Color Center in a Nanoscale Diamond Waveguide. *Phys. Rev. Lett.* **118**, 223603 (2017).

21. Becker, J. N. & Becher, C. Coherence Properties and Quantum Control of Silicon Vacancy Color Centers in Diamond (Phys. Status Solidi A 11∕2017). *physica status solidi (a)* **214**, 1770170 (2017).

22. C. T. Nguyen, D. D. Sukachev, M. K. Bhaskar, B. Machielse, D. S. Levonian, E. N. Knall, P. Stroganov, R. Riedinger, H. Park, M. Lončar, M. D. Lukin. Quantum network nodes based on diamond qubits with an efficient nanophotonic interface. *arXiv:1907.13199v2* (2019).

23. Siyushev, P. *et al.* Optical and microwave control of germanium-vacancy center spins in diamond. *Physical Review B* **96**, (2017).

24. Schröder, T. *et al.* Scalable focused ion beam creation of nearly lifetime-limited single quantum emitters in diamond nanostructures. *Nat. Commun.* **8**, 15376 (2017).

25. Mouradian, S., Wan, N. H., Schröder, T. & Englund, D. Rectangular photonic crystal nanobeam cavities in bulk diamond. *Appl. Phys. Lett.* **111**, 021103 (2017).

26. Wan, N. H., Mouradian, S. & Englund, D. Two-dimensional photonic crystal slab nanocavities on bulk single-crystal diamond. *Appl. Phys. Lett.* **112**, 141102 (2018).

27. Bernien, H. *et al.* Heralded entanglement between solid-state qubits separated by three metres. *Nature* **497**, 86–90 (2013).

28. Rogers, L. J. *et al.* Multiple intrinsically identical single-photon emitters in the solid state. *Nat. Commun.* **5**, 4739 (2014).

29. Meesala, S. *et al.* Strain engineering of the silicon-vacancy center in diamond. *Physical Review B* **97**, (2018).

30. Maity, S. *et al.* Spectral Alignment of Single-Photon Emitters in Diamond using Strain Gradient. *Physical Review Applied* **10**, (2018).

31. Machielse, B. *et al.* Quantum Interference of Electromechanically Stabilized Emitters in Nanophotonic Devices. *Physical Review X* **9**, (2019).

32. Carlo Bradac, Weibo Gao, Jacopo Forneris, Matt Trusheim, Igor Aharonovich. Quantum Nanophotonics with Group IV defects in Diamond. *arXiv:1906.10992* (2019).

33. Lodahl, P., Mahmoodian, S. & Stobbe, S. Interfacing single photons and single quantum dots with photonic nanostructures. *Reviews of Modern Physics* **87**, 347–400 (2015).

34. Zhong, T. *et al.* Optically Addressing Single Rare-Earth Ions in a Nanophotonic Cavity. *Physical Review Letters* **121**, (2018).

35. Dibos, A. M., Raha, M., Phenicie, C. M. & Thompson, J. D. Atomic Source of Single Photons in the Telecom Band. *Phys. Rev. Lett.* **120**, 243601 (2018).

36. Patra, B. *et al.* Cryo-CMOS Circuits and Systems for Quantum Computing Applications. *IEEE Journal of Solid-State Circuits* **53**, 309–321 (2018).

37. Kim, D. *et al.* A CMOS-integrated quantum sensor based on nitrogen–vacancy centres. *Nature Electronics* **2**, 284–289 (2019).

38. Stegmaier, M. & Pernice, W. H. P. Broadband directional coupling in aluminum nitride nanophotonic circuits. *Opt. Express* **21**, 7304–7315 (2013).

39. Xiong, C., Pernice, W. H. P. & Tang, H. X. Low-loss, silicon integrated, aluminum nitride photonic circuits and their use for electro-optic signal processing. *Nano Lett.* **12**, 3562–3568 (2012).

40. Zhu, S. & Lo, G.-Q. Aluminum nitride electro-optic phase shifter for backend integration on silicon. *Opt.*

*Express* **24**, 12501–12506 (2016).

41. Ghosh, S. & Piazza, G. Piezoelectric actuation of aluminum nitride contour mode optomechanical resonators. *Opt. Express* **23**, 15477–15490 (2015).

42. Wang, C. *et al.* Integrated lithium niobate electro-optic modulators operating at CMOS-compatible voltages. *Nature* **562**, 101–104 (2018).

43. Desiatov, B., Shams-Ansari, A., Zhang, M., Wang, C. & Lončar, M. Ultra-low-loss integrated visible photonics using thin-film lithium niobate. *Optica* **6**, 380 (2019).

44. Lo Piparo, N., Munro, W. J. & Nemoto, K. Quantum multiplexing. *Phys. Rev. A* **99**, 022337 (2019).

45. Bersin, E. *et al.* Individual control and readout of qubits in a sub-diffraction volume. *npj Quantum Information* **5**, (2019).

46. Pant, M., Choi, H., Guha, S. & Englund, D. Percolation based architecture for cluster state quantum computation using photon-mediated entanglement between atomic memories. *Preprint at https://arxiv.org/abs/1704. 07292* (2017).